\documentclass{article}
\usepackage{spconf,amsmath,graphicx,hyperref}
\usepackage{microtype}
\usepackage{multirow}
\usepackage{xcolor}

\title{SLAP: Scalable Language-Audio Pretraining with Variable-Duration Audio and Multi-Objective Training}
\name{\vspace{-6pt}\begin{tabular}{c}Xinhao Mei \qquad Gael Le Lan \qquad Haohe Liu \qquad Zhaoheng Ni \\ Varun Nagaraja \qquad Yang Liu \qquad Yangyang Shi \qquad Vikas Chandra \end{tabular}}
\address{Meta}

\begin{document}
\ninept
\maketitle
\begin{abstract}
Contrastive language-audio pretraining (CLAP) has achieved notable success in learning semantically rich audio representations and is widely adopted for various audio-related tasks. However, current CLAP models face several key limitations. First, they are typically trained on relatively small datasets, often comprising a few million audio samples. Second, existing CLAP models are restricted to short and fixed duration, which constrains their usage in real-world scenarios with variable-duration audio. Third, the standard contrastive training objective operates on global representations, which may hinder the learning of dense, fine-grained audio features. To address these challenges, we introduce Scalable Language-Audio Pretraining (SLAP), which scales language-audio pretraining to 109 million audio-text pairs with variable audio durations and incorporates multiple training objectives. SLAP unifies contrastive loss with additional self-supervised and captioning losses in a single-stage training, facilitating the learning of richer dense audio representations. The proposed SLAP model achieves new state-of-the-art performance on audio-text retrieval and zero-shot audio classification tasks, demonstrating its effectiveness across diverse benchmarks.
\end{abstract}
\section{Introduction}
\label{sec:intro}
Contrastive language-audio pretraining (CLAP) \cite{msclap, wu2023clap} has recently emerged as a powerful paradigm for learning semantically rich and transferable audio representations from paired audio-text data, inspired by the success of CLIP \cite{radford2021clip} in the vision domain. CLAP models have demonstrated strong performance across a variety of audio-related tasks, including zero-shot audio classification \cite{msclap}, audio captioning \cite{kim2024enclap}, and audio separation and generation \cite{liu2023audioldm, liu2025lass}. Despite these advances, current CLAP models face several critical limitations that hinder their effectiveness and scalability in real-world scenarios.

First, existing CLAP models are typically trained on relatively small datasets, often containing only a few million audio samples. In contrast, CLIP models benefit from billion-scale datasets, enabling them to learn more robust and generalizable features. Although recent efforts have leveraged large language models (LLMs) to generate captions for unlabeled audio \cite{mei2023wavcaps, audiosetcaps}, the scarcity of large-scale, high-quality audio-text pairs remains a significant bottleneck. As a consequence, most CLAP models initialize their audio encoders with weights from models pretrained on AudioSet \cite{audioset}, rather than training from scratch like CLIP models.

Second, most CLAP models with Transformer-based audio encoder are trained with short and fixed-duration audio clips, typically no longer than 10 seconds. In practice, shorter segments are padded to a predefined length, while longer segments are truncated, leading to inefficient computation and potential information loss. This duration limitation mainly arises from the reliance on audio encoders pretrained on AudioSet \cite{audioset} as well as the computational challenges associated with handling long and variable-length inputs. Although some recent approaches have attempted to extend CLAP models to longer and variable-length audio \cite{wu2023clap, feng2024elasticast}, these methods have not fully addressed this challenge.

Third, the contrastive training objective in CLAP focuses on aligning global representations of audio and text pairs. While this approach is effective for learning high-level semantic correspondences, it may limit the model’s ability to capture dense, fine-grained audio features that generalize well to downstream audio understanding tasks. Although multi-objective training with self-supervised and captioning objectives have been explored \cite{niizumi2025m2d2, cacophony}, they often introduce additional complexity with complex multi-stage training design.

To address these challenges, we propose Scalable Language-Audio Pretraining (SLAP), a new framework that improves contrastive language-audio pretraining along three dimensions. First, SLAP scales pretraining to 109 million audio-text pairs, significantly surpassing the scale of previous work. Second, SLAP natively supports variable audio durations by redesigning the Transformer-based audio encoder and training the encoder from scratch. Third, SLAP unifies contrastive, self-supervised and captioning losses into a single-stage training pipeline, facilitating the learning of richer and denser audio representations. SLAP achieves new state-of-the-art (SOTA) results on multiple audio-language benchmarks, highlighting the importance of large-scale, flexible, and multi-objective pretraining for audio representation learning.

In summary, our main contributions are:
\begin{itemize}
    \item We scale contrastive language-audio pretraining to 109 million audio-text pairs, an order of magnitude larger than previous work.
    \item We introduce a redesigned Transformer-based audio encoder that natively supports variable audio durations up to 30 seconds and is trained from scratch.
    \item We propose a unified, single-stage multi-objective training pipeline that combines contrastive, self-supervised, and captioning losses, enabling the learning of richer and denser audio representations.
    \item We achieve new state-of-the-art results on multiple audio-language benchmarks, demonstrating the effectiveness of our approach.
\end{itemize}

\section{Methods}
\label{sec:method}
Our proposed method introduces a redesigned audio Transformer encoder and the SLAP training framework. Section~\ref{ssec:audio_encoder} details the architecture and design choices of the audio encoder, while Section~\ref{ssec:training_method} presents the overall training pipeline.

\begin{figure}
    \centering
    \includegraphics[width=\linewidth]{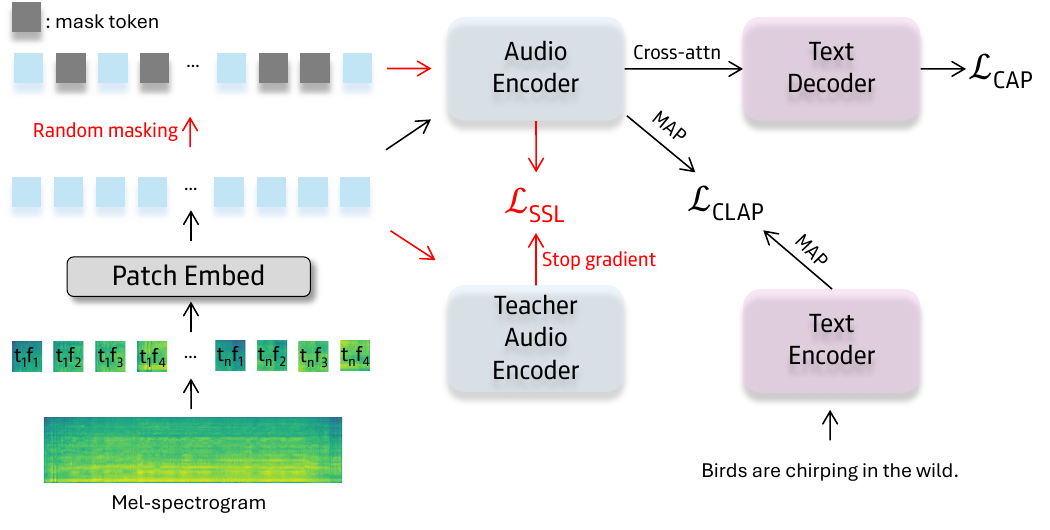}
    \caption{Overview of SLAP training framework, where the flow for self-supervised learning is annotated in red.}
    \label{fig:slap_framework}
\end{figure}

\subsection{Audio Encoder}
\label{ssec:audio_encoder}
The proposed audio encoder builds upon the Vision Transformer (ViT) architecture~\cite{vit}, but departs from prior work~\cite{wu2023clap} that directly transfers architectures and weights from pretrained vision models. Instead, we redesign the Transformer to better capture audio-specific characteristics and to natively support variable-length inputs.

\subsubsection{Modernized Transformer}
To enhance the standard ViT, we incorporate recent advancements from large language models~\cite{modernbert, Qwen2.5-VL}. Specifically, we employ 2D Rotary Positional Embedding (RoPE) to effectively model temporal and spectral relationships in the spectrograms. We adopt pre-normalization with RMSNorm and utilize the SwiGLU activation function to stabilize training \cite{Qwen2.5-VL}. Additionally, following~\cite{modernbert}, we remove bias terms from all linear layers.

Processing long audio sequences poses significant computational challenges, as attention scales quadratically with sequence length. To address this, we introduce an alternating attention mechanism that interleaves local sliding window attention with global attention. In local attention blocks, each patch token attends only to its neighbors within a small window, substantially reducing computational cost. Global attention blocks, in contrast, allow each token to attend to all others, enabling the modeling of long-range dependencies and global context. This hybrid design maintains efficiency for long audio inputs while preserving the encoder’s capacity to capture both fine-grained local patterns and global structures.

\subsubsection{Handling Variable Duration with Sequence Packing}

To efficiently support variable-duration audio clips during both training and inference, we adopt a sequence packing strategy. Each mel-spectrogram is first divided into non-overlapping patches, arranged in \textbf{time}$\rightarrow$\textbf{frequency} order. Although batching still requires padding waveforms to the maximum duration within a batch, our patch extraction ensures that all padded regions are grouped at the end of each sample. We then flatten and concatenate the patches from all audio clips in the batch, removing any padded patches, to form a single 1D packed sequence. This packed sequence is processed by the Transformer as a single long input with a batch size of one. To efficiently support this approach, we leverage Flash Attention~\cite{dao2023flashattn}, which natively handles attention over variable-length packed sequences and supports local sliding window attention. Compared to alternative methods~\cite{feng2024elasticast}, our approach simplifies the packing process and minimizes padding at the sample level, thereby enhancing computational efficiency and flexibility.

\subsection{Training Framework}
\label{ssec:training_method}
Given a collection of audio-text pairs $\{A_i, T_i\}$, where $A_i$ denotes an audio clip and $T_i$ its corresponding caption, our main objective is to develop a CLAP model that learns generalizable audio representations, suitable for a variety of single-modal and multimodal audio tasks. The proposed SLAP model consists of an audio encoder $f$, as described above, and a text encoder $g$. The audio encoder is trained from scratch, while the text encoder is initialized with weights from a pretrained ModernBERT~\cite{modernbert}. We adopt a unified, single-stage training framework that combines the standard CLAP loss with both a self-supervised learning loss and a captioning loss. An overview of the training framework is illustrated in Figure~\ref{fig:slap_framework}. The details of each objective are described below.

\subsubsection{Contrastive Loss} The contrastive training objective is a standard CLAP loss, which encourages embeddings of paired audio and text samples to be close, while pushing apart the embeddings of unpaired audio-text samples. To obtain global representations for both modalities, we use multi-head attention pooling (MAP) \cite{siglip} instead of conventional class token. Mathematically, this can be formulated as:
\begin{equation}
    E_i^a = \mathrm{MAP}_{\mathrm{audio}}(f(A_i))
\end{equation}
\begin{equation}
    E_i^t = \mathrm{MAP}_{\mathrm{text}}(g(T_i))
\end{equation}
where $E_i^a$ is the global audio embedding and $E_i^t$ is the global text embedding, both with a dimension of $D$. The CLAP loss is then defined as:
\begin{equation}
\resizebox{\columnwidth}{!}{$
\mathcal{L}_{\mathrm{CLAP}} = \frac{1}{2B} \sum_{i=1}^{B}\!\left(
    \log\frac{\exp(E_i^a \cdot E_i^t/\tau)}{\sum_{j=1}^B \exp(E_i^a \cdot E_j^t/\tau)}
  + \log\frac{\exp(E_i^t \cdot E_i^a/\tau)}{\sum_{j=1}^B \exp(E_i^t \cdot E_j^a/\tau)}
\right)
$}
\end{equation}
where $B$ is the batch size, and $\tau$ is a learnable temperature parameter.

\begin{table*}[!t]
 \vspace{-6pt}
  \caption{Experimental results of audio-text retrieval on test sets of AudioCaps and Clotho. Higher score means better performance.}
  \label{tab:retrieval_results}
  \centering
  \resizebox{0.85\textwidth}{!}{
  \begin{tabular}{c ccc | ccc | ccc | ccc}
    \hline
    \multirow{3}{*}{\textbf{Method}} & \multicolumn{6}{c}{\textbf{AudioCaps}} & \multicolumn{6}{c}{\textbf{Clotho}} \\
    \cline{2-13}  & 
    \multicolumn{3}{c}{\textbf{Text-to-Audio}} & \multicolumn{3}{c}{\textbf{Audio-to-Text}} & \multicolumn{3}{c}{\textbf{Text-to-Audio}} & \multicolumn{3}{c}{\textbf{Audio-to-Text}} \\
    \cline{2-13}
    & $\boldsymbol{R@1}$ & $\boldsymbol{R@5}$ & $\boldsymbol{R@10}$ & $\boldsymbol{R@1}$ & $\boldsymbol{R@5}$ & $\boldsymbol{R@10}$ & $\boldsymbol{R@1}$ & $\boldsymbol{R@5}$ & $\boldsymbol{R@10}$ & $\boldsymbol{R@1}$ & $\boldsymbol{R@5}$ & $\boldsymbol{R@10}$ \\
    \hline 
    \textbf{Zero-Shot} & & & & & & & & & & & & \\
    M2D2 \cite{niizumi2025m2d2} & 27.3 & 58.1 & 71.5 & 36.1 & 65.3 & 77.4 & 15.9 & 37.8 & 49.2 & 18.1 & 39.9 & 52.5 \\
    WavCaps \cite{mei2023wavcaps} & 28.6 & 61.1 & 75.8 & 40.2 & 69.4 & 80.3 & 16.5 & 38.8 & 50.9 & 20.0 & 43.3 & 56.6 \\
    SLAP (ours) & 35.1 & 65.8 & 76.4 & 44.8 & 73.9 & 84.4 & 18.9 & 41.7 & 54.3 & 23.5 & 46.6 & 59.0 \\
    \hline
    \textbf{Supervised} & & & & & & & & & & & & \\
    LAION-CLAP \cite{wu2023clap} & 36.2 & 70.3 & 82.5 & 45.0 & 76.7 & 88.0 & 17.2 & 42.9 & 55.4 & 24.2 & 51.1 & 66.9 \\
    WavCaps \cite{mei2023wavcaps} & 42.2 & 76.5 & 87.1 & 54.6 & 85.2 & 92.4 & 19.7 & 45.7 & 59.4 & 26.9 & 52.6 & 64.9 \\
    Cacophony \cite{cacophony} & 41.0 & 75.3 & 86.4 & 55.3 & 83.6 & 92.4 & 20.2 & 45.9 & 58.8 & 26.5 & 54.1 & 67.3 \\
    M2D2 \cite{niizumi2025m2d2}& 41.9 & 77.0 & 88.5 & 59.2 & 84.7 & 92.7 & 20.0 & 45.9 & 59.4 & 24.9 & 51.6 & 64.5 \\
    CED-LE \cite{cedle}  & 45.6 & \textbf{81.3} & \textbf{90.2} & 60.7 & 86.9 & \textbf{94.8} & 25.0 & 53.7 & 66.6 & 30.9 & 57.5 & 70.2 \\
    SLAP (ours) & \textbf{47.5} & 79.8 & 89.2 & \textbf{63.4} & \textbf{87.8} & 94.1 & \textbf{27.2} & \textbf{55.9} & \textbf{68.3} & \textbf{36.8} & \textbf{61.3} & \textbf{73.2} \\
    \hline
  \end{tabular}
  }
\end{table*}

\subsubsection{Self-Supervised Learning Loss} While the contrastive objective operates at the global embedding level, it does not directly encourage the learning of informative dense audio patch features. To address this, we introduce a masked audio modeling loss \cite{oquab2024dinov}. Specifically, we build a teacher audio encoder $f_t$ whose weights are updated through exponential moving average (EMA) of the main (student) audio encoder $f$. Before feeding patches into the Transformer blocks, a subset of patches is randomly replaced with a mask token $m$ according to a predefined masking ratio. The masked sequence is processed by the student encoder, while the unmasked sequence is processed by the teacher encoder. The encoded masked representations from the student encoder and their corresponding unmasked representations from the teacher encoder are projected onto prototype scores through a separate MLP head. Finally, a cross-entropy loss is computed between the softmax-normalized prototype scores of the student and teacher outputs. Formally, the self-supervised learning loss is defined as:
\begin{equation}
    \mathcal{L}_{\mathrm{SSL}} = - \sum_{i=1}^{B}\sum_{j}{Q_t}^{ij}\log{Q_s^{ij}}
\end{equation}
where $i$ is the sample index in a batch, $j$ are patch indices for masked tokens, $Q_t$ and $Q_s$ are teacher and student softmax-normalized prototype scores, respectively.

\subsubsection{Captioning Loss}
To further enrich the semantic content of the learned patch features, we incorporate a captioning loss. A shallow Transformer decoder is attached to generate captions conditioned on dense audio features before applying MAP head. The caption decoder is trained from scratch and optimized using standard teach-forcing training strategy, which can be formulated as:
\begin{equation}
    \mathcal{L}_{\mathrm{CAP}} = - \sum_{i=1}^{B}\sum_{z}P(T_{z}^i|T_{<z}^i, f(A_i))
\end{equation}
where $z$ denotes the index of a text token, $T_{z}^i$ is the $z$-th token in the $i$-th caption in a batch.

The overall training objective is a weighted combination of the three loss terms:
\begin{equation}
    \mathcal{L} = \alpha \mathcal{L}_{\mathrm{CLAP}} + \beta \mathcal{L}_{\mathrm{SSL}} + \gamma \mathcal{L}_{\mathrm{CAP}}
\end{equation}
where $\alpha$, $\beta$, and $\gamma$ are the weights for the respective loss terms.

\section{Experiments}
\label{sec:exp}

\subsection{Datasets and Evaluation Tasks}
We use the MovieGen Audio pretraining dataset~\cite{polyak2024moviegen} as our primary training corpus. This dataset contains approximately 109 million audio clips of varying durations, with captions generated by a general audio captioning model. Although the dataset itself is not publicly available, its curation pipeline can be readily applied to large-scale, unlabeled audio corpora.

To comprehensively assess our model's performance, we conduct evaluations across a diverse set of tasks on main audio-text benchmarks, including audio-text retrieval, zero-shot audio classification, audio captioning, and audio tagging. This evaluation protocol ensures a robust assessment of the model's generalization and transfer capabilities.

\subsection{Implementation Details}
Our audio encoder contains 12 Transformer layers, each with 12 attention heads, a hidden size of 768, and an intermediate dimension of 3072. We use an alternating attention mechanism, applying two sliding window attention blocks (window size of 24 frames) followed by a full attention block, enabling efficient modeling of both local and global dependencies. The teacher audio encoder is initialized from the student encoder and updated via exponential moving average at each training step. For text encoding, we initialize the encoder with a ModernBERT-base \cite{modernbert} model containing 22 Transformer layers, each with 12 attention heads and a hidden size of 768. The text decoder is trained from scratch and consists of 8 Transformer layers, each with 8 attention heads and a hidden size of 512.

Input waveforms are resampled to 16~kHz, and mel-spectrograms are computed using a 25~ms Hanning window, a 10~ms hop size, and 64 mel bins. The resulting spectrograms are partitioned into non-overlapping patches of size $16 \times 16$. We train the SLAP model on the MovieGen Audio dataset for 2 epochs with a batch size of 2048. The learning rate is linearly warmed up to $1 \times 10^{-4}$ over 2000 steps. SpecAugment is applied during training to improve robustness \cite{spec_augment}. The mask ratio is set to 0.5. The EMA momentum is warmed up from 0.994 to 1.0 using a cosine schedule. The loss weights $\alpha$, $\beta$, and $\gamma$ are empirically set to 1.0, 1.0, and 0.5, respectively.

\begin{table}[h]
\caption{Results of the top-1 accuracy on zero-shot audio classification.}
\centering
\resizebox{0.85\linewidth}{!}{
\begin{tabular}{c|cc|cc|c}
\hline
\textbf{Method}  & \textbf{ESC} &\textbf{US8K} & \textbf{CRD} & \textbf{RAD} & \textbf{GTZAN} \\
\hline
GLAP \cite{glap} & 88.8 & 78.9 & 20.5 & - & 69.6 \\
MS-CLAP \cite{msclap} & 93.9 & 82.3 & 30.0 & \textbf{31.5} & 58.4 \\ 
WavCaps \cite{mei2023wavcaps} & 94.8 & 80.6 & - & - & - \\
MGA-CLAP \cite{li2024mga_clap} & 94.9 &\textbf{83.7} & - & - & - \\
AudioSetCaps \cite{audiosetcaps} & 88.0 & 76.6 & 28.5 & 25.7 & 70.5 \\
M2D2 \cite{niizumi2025m2d2} & 94.3 & 82.9 & 28.6 & - & 79.3 \\
SLAP (ours) & 88.6 & 81.6 & 28.1 & 26.4 & 56.8 \\
SLAP$_\textrm{Wavcaps}$ (ours) & \textbf{95.5} & 83.5 & \textbf{32.2} & 29.8 & \textbf{80.5} \\
\hline
\end{tabular}
}
\label{tab:zs_ac_results}
\end{table}

\subsection{Results}

\noindent\textbf{Audio-Text Retrieval} We evaluate audio-text retrieval performance on the AudioCaps~\cite{audiocaps} and Clotho~\cite{clotho} datasets. AudioCaps consists of 10-second audio clips, while Clotho contains clips ranging from 15 to 30 seconds. We report both zero-shot and fine-tuned results. In the zero-shot setting, our SLAP model is evaluated directly on the target dataset without additional training. For fine-tuning, the model is further trained on each target dataset, retaining all three loss terms during this stage. Performance is measured using recall@$k$, and the main results are summarized in Table~\ref{tab:retrieval_results}.
In the zero-shot setting, SLAP achieves strong performance, approaching that of some supervised methods. Notably, although the captions in the MovieGen Audio pretraining dataset are generated by an audio captioning model, our results demonstrate the effectiveness of scaling up training data, even with noisy annotations. When fine-tuned on AudioCaps and Clotho, SLAP achieves state-of-the-art results, with recall@1 scores of 47.2\% and 63.4\% on AudioCaps, and 27.2\% and 36.8\% on Clotho, respectively, significantly outperforming previous approaches. We attribute these improvements not only to the increased scale of training data, but also to SLAP’s ability to process audio clips of variable duration, as evidenced by the results on Clotho. In contrast, prior methods require cropping audio to a fixed, short duration, which can lead to information loss.

\begin{table}[!t]
 \vspace{-6pt}
\caption{Audio captioning results on the test sets of AudioCaps and Clotho. A higher score means better performance.}
\label{table:caption_results}
\centering
\resizebox{0.9\linewidth}{!}{
\begin{tabular}[\linewidth]{ c | c | c c c } 
 \hline
 \textbf{Dataset} & \textbf{Method} & \textbf{METEOR} & \textbf{CIDEr} & \textbf{SPICE}\\ 
 \hline
 \multirow{5}{*}{Clotho}& {\color{gray}SLAM-AAC \cite{slam-aac}}  & {\color{gray} 19.7} & {\color{gray} 51.5} & {\color{gray} 14.8}\\
 & BLAT \cite{blat} & 17.8 & 41.5 & 12.6 \\ 
 & M2D2 \cite{niizumi2025m2d2} & 17.8 & 43.5 & 12.4 \\ 
 & Cacophony \cite{cacophony}&  15.3 & 41.5 & 10.6 \\ 
 & SLAP (ours) & \textbf{18.1} & \textbf{43.7} & \textbf{13.1}  \\ 
\hline
 \multirow{5}{*}{AudioCaps} & {\color{gray}AudioSetCaps \cite{audiosetcaps}} & {\color{gray} 26.2} & {\color{gray}83.9} & {\color{gray}18.6}\\
 & BLAT \cite{blat} & 24.7 & 73.3 & \textbf{18.4} \\ 
 & M2D2 \cite{niizumi2025m2d2} & 24.3 & 72.4 & 17.6 \\ 
 & Cacophony \cite{cacophony} & 23.6 & 72.8 & 16.8\\ 
 & SLAP (ours)& \textbf{24.9} & \textbf{75.1} & 18.1 \\ 
 \hline
\end{tabular}
}
\end{table}

\noindent\textbf{Zero-Shot Audio Classification} We evaluate zero-shot audio classification performance on five datasets spanning general sound (ESC-50 \cite{esc50}, UrbanSound8K \cite{us8k}), speech (CREMA-D \cite{crema-d}, RAVDESS \cite{ravdess}), and music (GTZAN \cite{gtzan}). In this setting, both input audio samples and class labels are encoded as audio and text embeddings, respectively. For each query, we compute the cosine similarity between the audio embedding and each class label embedding, assigning the class with the highest similarity as the predicted label. Since the MovieGen Audio captions are generated by an audio captioning model, they may introduce noise, hallucination and fail to cover the full diversity of sound events. To address this, we further fine-tune the SLAP model on the WavCaps~\cite{mei2023wavcaps} dataset, which provides broader coverage of real-world sound events. We report results for both pretraining and fine-tuning settings. Table~\ref{tab:zs_ac_results} presents the top-1 accuracy for each dataset. Our SLAP model, when fine-tuned on WavCaps, achieves new state-of-the-art results on ESC-50 (95.5\%), CREMA-D (32.2\%), and GTZAN (80.5\%), and remains competitive on UrbanSound8K and RAVDESS. The performance gains from fine-tuning suggest that supplementing generated captions with additional weakly-labeled data helps improve the model's ability to generalize across diverse audio domains. These results highlight the effectiveness of SLAP and the importance of scaling up training data.

\noindent\textbf{Audio Captioning} We evaluate audio captioning performance on the AudioCaps and Clotho datasets. For these experiments, we use the same fine-tuned checkpoints as in the audio-text retrieval task, retaining all three loss terms. Notably, the captioning loss is assigned a lower weight and serves primarily as an auxiliary objective. We report results using METEOR, CIDEr, and SPICE, as shown in Table~\ref{table:caption_results}. For fair comparison, we primarily benchmark against other CLAP-based methods. SLAM-AAC~\cite{slam-aac} and AudioSetCaps~\cite{audiosetcaps}, shown in gray, represent SOTA models on Clotho and AudioCaps, respectively. Despite using the captioning loss only as an auxiliary objective, our model outperforms M2D2~\cite{niizumi2025m2d2} and Cacophony~\cite{cacophony} on both AudioCaps and Clotho. Both M2D2 and Cacophony employ multi-stage training frameworks that combine CLAP and self-supervised objectives, with Cacophony also incorporating a captioning loss. These results demonstrate the effectiveness of our unified training pipeline, which significantly simplifies the process compared to multi-stage approaches. Compared to SOTA captioning models, which typically leverage powerful pretrained large language models as text decoders \cite{slam-aac}, SLAP uses a shallow text decoder trained from scratch. This difference, along with the lower weight assigned to the captioning loss, may contribute to the performance gap between SLAP and the best-performing captioning methods.

\noindent\textbf{Audio Tagging} To further evaluate the effectiveness of our audio encoder, we conduct audio tagging experiments on several benchmark datasets, including AudioSet, ESC-50, and Speech Commands V2 (SPC-2)~\cite{niizumi2025m2d2}. Specifically, we attach a linear classification layer to the audio encoder of the pretrained SLAP model and fine-tune the entire audio encoder on each target dataset. For AudioSet, we follow the fine-tuning protocol from AudioMAE~\cite{huang2022amae}, incorporating balanced sampling, mixup, and cyclic rolling augmentation. Table~\ref{tab:tagging_results} reports mean average precision (mAP) on AudioSet and classification accuracy on the other datasets. We compare our approach against other CLAP-based methods, while current SOTA (non-CLAP-based) results are also included for reference. The results show that SLAP outperforms BLAT~\cite{blat} but falls short of M2D2~\cite{niizumi2025m2d2} on AudioSet. This performance gap may be due to M2D2 did self-supervised training on AudioSet, whereas we only apply direct audio tagging fine-tuning. On SPC-2, SLAP achieves the highest score, and on ESC-50, it shows competitive results compared to M2D2. These findings highlight the effectiveness and generalizability of our audio encoder and training pipeline. 

\begin{table}[!t]
 \vspace{-6pt}
\caption{Results of audio tagging among CLAP-based methods.}
\centering
\resizebox{0.7\linewidth}{!}{
\begin{tabular}{c|ccc}
\hline
\textbf{Model}  & \textbf{Audioset} &\textbf{ESC-50} & \textbf{SPC-2} \\
\hline
\color{gray}{SOTA} & \color{gray}{50.0 \cite{ced}} & \color{gray}{98.1 \cite{beats}} &  \color{gray}{98.7 \cite{xin2023masked}} \\
BLAT \cite{blat} & 44.0 & 95.8  & - \\
MS-CLAP \cite{msclap}& - & 96.7 &  96.8 \\
M2D2 \cite{niizumi2025m2d2} & \textbf{49.0} & \textbf{98.5} &  98.4 \\
SLAP (ours) & 47.8 & 98.2 & \textbf{98.5} \\
\hline
\end{tabular}
}
\label{tab:tagging_results}
\end{table}

\begin{table}[!h]
  \caption{Ablation results of audio-text retrieval on  AudioCaps.}
  \label{tab:ablation_results}
  \centering
  \resizebox{0.95\linewidth}{!}{
  \begin{tabular}{c ccc | ccc }
    \hline
    \multirow{2}{*}{\textbf{Method}} 
    & \multicolumn{3}{c}{\textbf{Text-to-Audio}} 
    & \multicolumn{3}{c}{\textbf{Audio-to-Text}}  \\
    \cline{2-7}
    & $\boldsymbol{R@1}$ & $\boldsymbol{R@5}$ & $\boldsymbol{R@10}$ & $\boldsymbol{R@1}$ & $\boldsymbol{R@5}$ & $\boldsymbol{R@10}$ \\
    \hline 
    SLAP  & 35.1 & 65.8 & 79.4 & 44.8 & 73.9 & 84.4 \\
     \quad w/o $\mathcal{L}_{\mathrm{SSL}}$ & 32.8 & 64.4 & 77.4 & 42.4 & 72.4 & 82.2  \\
     \quad w/o $\mathcal{L}_{\mathrm{CAP}}$ &  34.7 & 65.2 & 78.5 & 43.0 & 73.4 & 83.7 \\
     \quad w/o $\mathcal{L}_{\mathrm{SSL}}$ \& $\mathcal{L}_{\mathrm{CAP}}$ & 32.6 & 63.0 & 75.9  & 41.7 & 73.3 & 83.5 \\
     \quad w/o local attn & 34.2 & 65.4 & 78.0 & 43.4 & 73.4 & 84.1 \\
    \hline
  \end{tabular}
  }
\end{table}

\subsection{Ablation Study}
We perform ablation studies to evaluate the contributions of individual loss terms and the impact of local window attention. Pretrained models are assessed on the AudioCaps dataset in the zero-shot audio-text retrieval setting. Table~\ref{tab:ablation_results} presents the results. The results indicate that both the self-supervised learning loss and the captioning loss contribute to performance improvements, with the self-supervised learning loss having a more substantial impact and the captioning loss providing a modest benefit. Furthermore, incorporating local window attention leads to better results than using full attention alone.

\section{Conclusion}
\label{sec:conclu}
In this work, we propose scalable language-audio pretraining (SLAP), a novel framework that addresses key limitations of existing CLAP models by scaling pretraining to 109 million audio-text pairs, supporting variable-length audio inputs, and unifying multiple training objectives in a single-stage pipeline. Extensive experiments demonstrate that SLAP achieves state-of-the-art performance across multiple audio-language benchmarks.

\vfill\pagebreak

\bibliographystyle{IEEEbib}
\bibliography{strings,refs}

\begin{thebibliography}{10}

\bibitem{msclap}
Benjamin Elizalde, Soham Deshmukh, Mahmoud~Al Ismail, and Huaming Wang,
\newblock ``{CLAP} learning audio concepts from natural language supervision,''
\newblock in {\em ICASSP}, 2023.

\bibitem{wu2023clap}
Yusong Wu, Ke~Chen, Tianyu Zhang, Yuchen Hui, Taylor Berg-Kirkpatrick, et~al.,
\newblock ``Large-scale contrastive language-audio pretraining with feature fusion and keyword-to-caption augmentation,''
\newblock in {\em ICASSP}, 2023.

\bibitem{radford2021clip}
Alec Radford, Jong~Wook Kim, Chris Hallacy, A.~Ramesh, Gabriel Goh, et~al.,
\newblock ``Learning transferable visual models from natural language supervision,''
\newblock in {\em ICML}, 2021.

\bibitem{kim2024enclap}
Jaeyeon Kim, Jaeyoon Jung, Jinjoo Lee, and Sang~Hoon Woo,
\newblock ``{EnCLAP}: Combining neural audio codec and audio-text joint embedding for automated audio captioning,''
\newblock in {\em ICASSP}, 2024.

\bibitem{liu2023audioldm}
Haohe Liu, Zehua Chen, Yi~Yuan, Xinhao Mei, Xubo Liu, et~al.,
\newblock ``{AudioLDM}: Text-to-audio generation with latent diffusion models,''
\newblock {\em ICML}, 2023.

\bibitem{liu2025lass}
Xubo Liu, Qiuqiang Kong, Yan Zhao, Haohe Liu, Yi~Yuan, et~al.,
\newblock ``Separate anything you describe,''
\newblock {\em TASLP}, 2025.

\bibitem{mei2023wavcaps}
Xinhao Mei, Chutong Meng, Haohe Liu, Qiuqiang Kong, Tom Ko, et~al.,
\newblock ``Wav{C}aps: A {ChatGPT}-assisted weakly-labelled audio captioning dataset for audio-language multimodal research,''
\newblock {\em TASLP}, 2024.

\bibitem{audiosetcaps}
Jisheng Bai, Haohe Liu, Mou Wang, Dongyuan Shi, Wenwu Wang, et~al.,
\newblock ``{AudioSetCaps}: An enriched audio-caption dataset using automated generation pipeline with large audio and language models,''
\newblock {\em TASLP}, 2025.

\bibitem{audioset}
Jort~F. Gemmeke, Daniel P.~W. Ellis, Dylan Freedman, Aren Jansen, Wade Lawrence, et~al.,
\newblock ``Audio{S}et: An ontology and human-labeled dataset for audio events,''
\newblock in {\em ICASSP}, 2017.

\bibitem{feng2024elasticast}
Jiu Feng, Mehmet~Hamza Erol, Joon~Son Chung, and Arda Senocak,
\newblock ``Elastic{AST}: An audio spectrogram {T}ransformer for all length and resolutions,''
\newblock in {\em Interspeech}, 2024.

\bibitem{niizumi2025m2d2}
Daisuke Niizumi, Daiki Takeuchi, Masahiro Yasuda, Binh~Thien Nguyen, Yasunori Ohishi, et~al.,
\newblock ``{M2D2}: Exploring general-purpose audio-language representations beyond {CLAP},'' 2025.

\bibitem{cacophony}
Ge~Zhu, Jordan Darefsky, and Zhiyao Duan,
\newblock ``Cacophony: An improved contrastive audio-text model,''
\newblock {\em TASLP}, 2024.

\bibitem{vit}
Alexey Dosovitskiy, Lucas Beyer, Alexander Kolesnikov, Dirk Weissenborn, Xiaohua Zhai, et~al.,
\newblock ``An image is worth 16x16 words: {T}ransformers for image recognition at scale,''
\newblock in {\em ICLR}, 2021.

\bibitem{modernbert}
Benjamin Warner, Antoine Chaffin, Benjamin Clavié, Orion Weller, Oskar Hallström, et~al.,
\newblock ``Smarter, better, faster, longer: A modern bidirectional encoder for fast, memory efficient, and long context finetuning and inference,'' 2024.

\bibitem{Qwen2.5-VL}
Shuai Bai, Keqin Chen, Xuejing Liu, Jialin Wang, Wenbin Ge, et~al.,
\newblock ``Qwen2.5-{VL} technical report,''
\newblock {\em arXiv preprint arXiv:2502.13923}, 2025.

\bibitem{dao2023flashattn}
Tri Dao,
\newblock ``Flash{A}ttention-2: Faster attention with better parallelism and work partitioning,''
\newblock in {\em ICLR}, 2024.

\bibitem{siglip}
Xiaohua Zhai, Basil Mustafa, Alexander Kolesnikov, and Lucas Beyer,
\newblock ``Sigmoid loss for language image pre-training,''
\newblock in {\em ICCV}, 2023.

\bibitem{cedle}
Zhiyong Yan, Heinrich Dinkel, Yongqing Wang, Jizhong Liu, Junbo Zhang, et~al.,
\newblock ``Bridging language gaps in audio-text retrieval,''
\newblock in {\em Interspeech}, 2024.

\bibitem{oquab2024dinov}
Maxime Oquab, Timoth{\'e}e Darcet, Th{\'e}o Moutakanni, Huy~V. Vo, Marc Szafraniec, et~al.,
\newblock ``{DINO}v2: Learning robust visual features without supervision,''
\newblock {\em TMLR}, 2024.

\bibitem{polyak2024moviegen}
Adam Polyak, Amit Zohar, Andrew Brown, Andros Tjandra, Animesh Sinha, et~al.,
\newblock ``Movie {G}en: A cast of media foundation models,''
\newblock {\em arXiv preprint arXiv:2410.13720}, 2024.

\bibitem{spec_augment}
Daniel~S. Park, William Chan, Yu~Zhang, Chung-Cheng Chiu, Barret Zoph, et~al.,
\newblock ``Spec{A}ugment: A simple data augmentation method for automatic speech recognition,''
\newblock {\em Interspeech}, 2019.

\bibitem{glap}
Heinrich Dinkel, Zhiyong Yan, Tianzi Wang, Yongqing Wang, Xingwei Sun, et~al.,
\newblock ``{GLAP}: General contrastive audio-text pretraining across domains and languages,'' 2025.

\bibitem{li2024mga_clap}
Yiming Li, Zhifang Guo, Xiangdong Wang, and Hong Liu,
\newblock ``Advancing multi-grained alignment for contrastive language-audio pre-training,''
\newblock in {\em ACM Multimedia}, 2024.

\bibitem{audiocaps}
Chris~Dongjoo Kim, Byeongchang Kim, Hyunmin Lee, and Gunhee Kim,
\newblock ``{AudioCaps: Generating Captions for Audios in The Wild},''
\newblock in {\em NAACL-HLT}, 2019.

\bibitem{clotho}
Konstantinos Drossos, Samuel Lipping, and Tuomas Virtanen,
\newblock ``Clotho: an audio captioning dataset,''
\newblock in {\em ICASSP}, 2020.

\bibitem{slam-aac}
Wenxi Chen, Ziyang Ma, Xiquan Li, Xuenan Xu, Yuzhe Liang, et~al.,
\newblock ``Slam-aac: Enhancing audio captioning with paraphrasing augmentation and clap-refine through llms,''
\newblock in {\em ICASSP}, 2025.

\bibitem{blat}
Xuenan Xu, Zhiling Zhang, Zelin Zhou, Pingyue Zhang, Zeyu Xie, et~al.,
\newblock ``{BLAT}: Bootstrapping language-audio pre-training based on audioset tag-guided synthetic data,''
\newblock in {\em ACM Multimedia}, 2023.

\bibitem{esc50}
Karol~J. Piczak,
\newblock ``{ESC}: {Dataset} for {Environmental Sound Classification},''
\newblock in {\em ACM {Multimedia}}, 2015.

\bibitem{us8k}
Justin Salamon, Christopher Jacoby, and Juan~Pablo Bello,
\newblock ``A dataset and taxonomy for urban sound research,''
\newblock in {\em ACM Multimedia}, 2014.

\bibitem{crema-d}
Houwei Cao, David~G. Cooper, Michael~K. Keutmann, Ruben~C. Gur, Ani Nenkova, et~al.,
\newblock ``{CREMA-D}: Crowd-sourced emotional multimodal actors dataset,''
\newblock {\em TAC}, 2014.

\bibitem{ravdess}
Steven~R Livingstone and Frank~A Russo,
\newblock ``The ryerson audio-visual database of emotional speech and song ({RAVDESS}): A dynamic, multimodal set of facial and vocal expressions in north american english,''
\newblock {\em PloS one}, 2018.

\bibitem{gtzan}
G.~Tzanetakis and P.~Cook,
\newblock ``Musical genre classification of audio signals,''
\newblock {\em TASLP}, 2002.

\bibitem{huang2022amae}
Po-Yao Huang, Hu~Xu, Juncheng Li, Alexei Baevski, Michael Auli, et~al.,
\newblock ``Masked autoencoders that listen,''
\newblock in {\em NeurIPS}, 2022.

\bibitem{ced}
Heinrich Dinkel, Yongqing Wang, Zhiyong Yan, Junbo Zhang, and Yujun Wang,
\newblock ``{CED}: Consistent ensemble distillation for audio tagging,''
\newblock in {\em ICASSP}, 2024.

\bibitem{beats}
Sanyuan Chen, Yu~Wu, Chengyi Wang, Shujie Liu, Daniel Tompkins, et~al.,
\newblock ``{BEAT}s: Audio pre-training with acoustic tokenizers,''
\newblock in {\em ICML}, 2023, pp. 5178--5193.

\bibitem{xin2023masked}
Yifei Xin, Xiulian Peng, and Yan Lu,
\newblock ``Masked audio modeling with {CLAP} and multi-objective learning,''
\newblock in {\em Interspeech}, 2023.

\end{thebibliography}

\end{document}